\documentclass[a4paper,12pt]{article}
\usepackage{latexsym}

\begin{document}

\title{
\begin{flushright}
{{\small {DFPD 00/TH/44
\vskip-0.5cm
 hep-th/0011020}}}
\end{flushright}
~\\
{\bf Geometrical aspects of superbrane dynamics}
}

\bigskip
\author{Paolo Pasti$^{1}$, Dmitri Sorokin$^{1,2}$ and Mario
Tonin$^{1}$\\
~\\
{$~^1$ \it Universit\`a Degli Studi di Padova,
Dipartimento di Fisica ``Galileo Galilei'' ed}\\
{\it INFN, Sezione di Padova Via F. Marzolo, 8, 35131 Padova,
Italia}\\
~\\
{$~^2$ \it Institute for Theoretical Physics}\\
{\it NSC ``Kharkov Institute of Physics and Technology''}\\
{\it 61108 Kharkov, Ukraine} }
\date{}


\maketitle
\thispagestyle{empty}


\begin{abstract}
The geometrical (superembedding) approach is used as a tool for
deriving from the worldvolume dynamics of superbranes field
theoretical models exhibiting partial supersymmetry breaking. In
this way we obtain nonlinear actions for Goldstone superfields
associated with physical degrees of freedom of the superbranes.
\end{abstract}

\newpage
\section{Introduction}
In this contribution we discuss recent results in the construction
of superfield actions for effective field theories on superbrane
worldvolumes. These effective supersymmetric field theories
describe (in a static or physical gauge) superbrane fluctuations
in a supergravity background and the worldvolume dynamics of pure
brane fields, such as vector gauge fields of the Dirichlet branes.

The present interest in the field theories associated with branes
is at least threefold:
\begin{itemize}
\item
these are models which exhibit (partial) breaking of
supersymmetry, i.e. a property which is important for
phenomenological applications of supersymmetry;
\item
in the case of Anti--de--Sitter backgrounds these effective
theories are associated with superconformal field theories on the
AdS boundary, and thus are used to test the AdS/CFT correspondence
conjecture;
\item
the knowledge of an explicit form of brane effective actions
should be useful for the development of various brane world
sceanrii, which are under intensive study these days.
\end{itemize}

In this paper we will concentrate on the consideration of the
first of the items above, namely, on the relationship of
superbrane worldvolume actions with models of partial breaking of
global supersymmetries. The mechanism of partial supersymmetry
breaking caused by superbrane solutions has been under study over
a long period of time starting from References \cite{polch,achu}.

Recently, superfield actions for Goldstone supermultiplets which
cause 1/2 supersymmetry breaking in $D=4$ and $D=3$ theories have
been constructed in
\cite{bg2,rt,ik1}.

For instance, it has been shown \cite{bg2} that spontaneous
breaking of global $N=2$ supersymmetry down to $N=1$ in $D=4$
caused by an
$N=1$ vector supermultiplet is described by the action for a
Goldstone--vector supermultiplet which produces the
Dirac--Born--Infeld action in the bosonic sector of the theory
\cite{fer}. This implies that the supersymmetric model thus obtained
should have something to do with a space--filling Dirichlet
3--brane. However, the relation between the fermionic sector of
this model and the fermionic fluctuations of the D3--brane
(described by a Green--Schwarz--like action \cite{d3}) has not
been established, since the methods used for the construction of
the Goldstone superfield actions differ from the methods used for
the construction of superbrane actions. The former are based on
methods of non--linear realizations of spontaneously broken
symmetries.

The same situation is in a simpler case of breaking $N=2$
supersymmetry down to $N=1$ in $D=3$ by a Goldstone scalar
supermultiplet \cite{ik1} which is believed to be associated with
supermembrane fluctuations in an $N=1$, $D=4$ superspace.

In what follows we propose a generic procedure of how one can
arrive at the Goldstone superfield actions starting directly from
corresponding covariant actions for superbranes. To reach this
goal we shall use the superembedding approach  which is a generic
geometrical method for describing super--p--branes, Dp-branes and
M--branes (see \cite{s} for a review and references).

To be concrete we shall describe the basic properties of the
superembedding approach and the derivation of Goldstone
superfield actions exhibiting partial supersymmetry breaking with
the example of a supermembrane propagating in an $N=1$, $D=4$
target superspace, though we should note that the same reasoning
is also applicable to more physically interesting, but more
complicated, cases of other branes, such as a D3--brane.

\section{The $N=1$, $D=4$ supermembrane}
Our strategy will be
\begin{description}
\item[i)]
to define the conditions of superembedding describing
supermembrane dynamics;
\item[ii)]
to construct a worldvolume superfield action for the $N=1$, $D=4$
supermembrane;
\item[iii)]
to gauge fix all local worldvolume symmetries of the action by
imposing a physical gauge and to solve for the superembedding
condition in terms of an independent (Goldstone) superfield;
\item[iv)]
to reduce the supermembrane action to a Goldstone superfield
action with $N=2$, $d=3$ supersymmetry broken down to $N=1$,
$d=3$.
\end{description}

\subsection{The superembedding condition}
For superembedding to be relevant to the description of the
dynamics of superbranes one should find an appropriate condition
of how a worldvolume supersurface is embedded into target
superspace. The basic superembedding condition has a very simple
geometrical meaning and is generic for all types of superbranes.

Consider a worldvolume supersurface ${\cal M}$ parametrized by
$d=3$ bosonic coordinates $\xi^m$ and ${2}$ fermionic
coordinates $\eta^\mu$, which we will collectively call
\begin{equation}\label{z}
z^M=(\xi^m,~\eta^\mu), \quad m=0,1,2, \quad \mu=1,2.
\end{equation}

The geometry of ${\cal M}$ is described, in a superdiffeomorphism
invariant way, by a set of supervielbein one--forms
\begin{equation}\label{e}
e^A(z)=dz^Me^{~A}_M=\left(e^a(\xi,\eta),~e^\alpha(\xi,\eta)\right),
\end{equation}
which form a local basis in the cotangent space of ${\cal M}$. The
indices $a$ and $\alpha$ are, respectively, the indices of the
vector and a spinor representation of the group $SO(1,2)$ of local
rotations in the cotangent space.

We would like to embed this supersurface into a curved target
superspace
$\underline M$ parametrized by $D=4$ bosonic coordinates
$X^{\underline m}$ and {\bf 4} fermionic coordinates
$\Theta^{\underline\mu}$, which we will collectively call
\begin{equation}\label{Z}
Z^{\underline M}=(X^{\underline m},~\Theta^{\underline\mu}), \quad
\underline m=0,\dots,3, \quad \underline\mu=1,\dots,4.
\end{equation}
Note that for embedding we have chosen a supersurface with the
number of Grassmann--odd directions being half the number of
target--superspace Grass\-mann--odd directions. This is for being
able to identify ${\bf 2}$ local worldvolume supersymmetries with
${\bf 2}$ independent fermionic $\kappa$--symmetries of the standard
(Green--Schwarz) formulation of supermembrane dynamics by
Bergshoeff, Sezgin and Townsend \cite{bst1}.

The geometry of the target superspace is described in a
superdiffeomorphism invariant way by a set of supervielbein
one--forms
\begin{equation}\label{E}
E^{\underline A}(Z)=dZ^{\underline M}E^{~\underline A}_{\underline
M}=\left( E^{\underline a}(X,\Theta), E^{\underline
\alpha}(X,\Theta)\right),
\end{equation}
which form a local frame in the cotangent space of the target
superspace. The indices $\underline a$ and $\underline\alpha$ are,
respectively, the indices of the vector and a spinor
representation of the group $SO(1,D-1)$ of local rotations in the
$\underline M$ cotangent space.

Superembedding is a map of ${\cal M}$ into $\underline M$ which is
locally described by $X^{\underline m}$ and
$\Theta^{\underline\mu}$ as functions of the supersurface
coordinates
\begin{equation}\label{map}
z^M\quad \rightarrow \quad Z^{\underline M}(z)=\left(X^{\underline
m}(\xi, \eta),~\Theta^{\underline\mu}(\xi,\eta)\right).
\end{equation}
 The map induces the pullback onto the supersurface of the
target superspace one--form (\ref{E}). In particular, the vector
supervielbein $E^{\underline a}$ pullback is a one--superform on
the supersurface. It has the following decomposition in the local
basis (\ref{e}) on $\cal M$
\begin{equation}\label{pull}
E^{\underline a}(z)=e^a(z)E^{~\underline
a}_a(Z(z))+e^\alpha(z)E^{~\underline a}_\alpha(Z(z)).
\end{equation}
The superembedding condition we are interested in is the
vanishing of the worldvolume spinor components of $E^{\underline
a}(z)$
\begin{equation}\label{emb}
E^{~\underline a}_\alpha(Z(z))=0.
\end{equation}
In other words eq. (\ref{emb}) is a superfield constraint on
(\ref{map}) which singles out the superembeddings such that the
pullback of the supervielbein $E^{\underline a}$ has non--zero
components only along vector directions of the supersurface.

In addition we assume that the supersurface and
target--superspace geometry satisfy torsion constraints of
corresponding $d=3$ and $D=4$ supergravity
\begin{equation}\label{torsion}
T^a=-ie^\alpha e^\beta \gamma_{\alpha\beta}^a +
\cdots, \quad T^{\underline a}=
-iE^{\underline\alpha} E^{\underline\beta}
\Gamma_{\underline\alpha\underline\beta}^{\underline a} +
\cdots,
\end{equation}
which are consistent with the superembedding condition
($\gamma^a$ and $\Gamma^{\underline a}$ are, respectively, $d=3$
and $D=4$ Dirac matrices). So, we deal with two supergravity
theories embedded one into another. What is the role of the
superembedding condition (\ref{emb}) then?

In some cases the superembedding condition produces only
``kinematic'' constraints (e.g. the Virasoro constraints for
superstrings) and does not put superbrane dynamics on the mass
shell. In these cases (as the $N=1$, $D=4$ supermembrane
considered here) worldvolume superfield actions can be
constructed.

In other cases the superembedding condition contains all the
constraints and dynamical equations of motion of corresponding
superbranes (e.g. a D=11 supermembrane \cite{bpstv} and an
M5--brane \cite{hs}).

In all the cases the superembedding condition ensures that the
superworldvolume geometry is induced by the embedding
\footnote{In particular, it can be shown that the worldvolume
metric is an induced metric defined in a standard way as
\begin{equation}\label{indmet}
g_{mn}(\xi)=\partial_mZ^{\underline M}E_{\underline M}^{\underline
a}\partial_nZ^{\underline N} E_{\underline N}^{\underline
b}\eta_{\underline{ab}}|_{\eta=0}.
\end{equation}},
i.e. that there is no fully fledged propagating supergravity on
the brane worldvolume.

\subsection{The superembedding action}
To construct the supermembrane action we should introduce one
more notion, namely, a target--superspace three--form gauge
superfield $A_{\underline{NML}}(X,\Theta)$, whose field strength
obeys the constraint\footnote{Note that in $D=4$, since the dual
field strength
${}^*F^{(4)}=const$ on the mass shell, $A^{(3)}$ does not have physical
degrees of freedom, but its vacuum energy may contribute to the
value of the cosmological constant.}
\begin{equation}\label{7}
F^{(4)}=dA^{(3)}={i\over 2}E^{\underline a}E^{\underline b}\bar
E_{\underline\alpha}E^{\underline\beta} (\Gamma_{\underline a
\underline b})^{\underline
\alpha}_
{~\underline\beta} +{1\over{4!}}E^{\underline a}E^{\underline b}
E^{\underline c}E^{\underline d} F_{\underline a\underline
b\underline c\underline d}.
\end{equation}
It is well known that the supermembrane minimally couples to this
gauge superfield, which in the Green--Schwarz formulation
\cite{bst1} is described by the Wess--Zumino term
\begin{equation}\label{wz}
S_{WZ}={T\over 2}\int d^3\xi \varepsilon^{mnp}A_{mnp}(X,\Theta),
\end{equation}
where $T$ is the membrane tension and
$A_{mnp}\left(X(\xi),\Theta(\xi)\right)=\partial_mZ^{\underline M}
\partial_nZ^{\underline N}\partial_pZ^{\underline P}A_{\underline{PNM}}$
is the worldvolume pullback of
$A^{(3)}$. In the superembedding formulation the $A^{(3)}$
pullback is a worldvolume three superform
\begin{equation}\label{a3pull}
A^{(3)}|_{\cal M}={1\over{3!}}e^A(z)e^B(z)e^C(z)~A_{CBA},
\end{equation}
and the supermembrane action is constructed as an integral over
the supersurface ${\cal M}~ (z^M=(\xi^m,\eta^\mu))$. It can be
checked that because of dimensional reasons the only component of
the $A^{(3)}$ pullback which can enter the action is the one with
two spinor and one vector indices $A_{\alpha\beta a}$. Thus, we
assume the supermembrane action in the superembedding approach to
have the following form \cite{pst} (it is a ``brany"
generalization of an $N=1$ superstring action proposed in
\cite{tonin})
\begin{equation}\label{a2}
S={1\over{3!}}\int d^3\xi
d^2\eta~sdet{~e}~\gamma^{a\alpha\beta}A_{\alpha\beta a}+
\int d^3\xi d^2\eta P^\alpha_{\underline a}E^{~\underline
a}_{\alpha} \,,
\end{equation}
whose second Lagrange multiplier term takes care of the
superembedding condition, and $sdet{~e}$ is the superdeterminant
of the worldvolume supervielbein matrix $e^{~A}_M(z)$ (\ref{e}).

Integrating over the Grassmann--odd variables and eliminating
auxiliary fields with the use of the superembedding condition,
one can check that this action reduces to the conventional
supermembrane action \cite{bst1}.

The action (\ref{a2}) is invariant under the following symmetries:
\begin{description}
\item[i)]
worldvolume and target space superdiffeomorphisms
\begin{equation}\label{sd}
z^M~\rightarrow~f^M(\xi,\eta), \quad Z^{\underline
M}~\rightarrow~f^{\underline M}(X,\Theta),
\end{equation}
\item[ii)]
super--Weyl transformations
\begin{equation}\label{superW}
{e'}^a=W^2(z)e^a, \quad
{e'}^\alpha=W(z)e^\alpha-ie^a\gamma_a^{\alpha\beta}{\cal D}_\beta
W,
\end{equation}
\item[iii)]
local $SO(1,2)$ rotations in the tangent space of the
superworldvolume.
\end{description}

The action (\ref{a2}) describes the supermembrane in an arbitrary
$N=1$, $D=4$ supergravity background\footnote{Actually, the action (\ref{a2})
also describes supermembranes in $D=5,7$ and 11 superspaces
\cite{pst}.}, but since here we are interested in effects of
spontaneous breaking of global supersymmetry, let us choose the
superbackground to be flat. Then the supervielbeins and the gauge
superfield take the form
\begin{equation}\label{calE}
{E}^{\underline a}={dX}^{\underline
a}-id\bar\Theta\Gamma^{\underline{a}}\Theta, \quad
E^{\underline\alpha}=d\Theta^{\underline\alpha},
\end{equation}
\begin{equation}\label{Aflat}
A^{(3)}=i\bar\Theta\Gamma_{\underline{ab}}d\Theta({
E}^{\underline a}{E}^{\underline b}-i{E}^{\underline a}
\bar\Theta\Gamma^{\underline{b}}d\Theta-
{1\over
3}\bar\Theta\Gamma^{\underline{a}}d\Theta\bar\Theta\Gamma^{\underline{b}}d\Theta)
\end{equation}
and the superembedding condition is
\begin{equation}\label{embflat}
{E}^{~\underline a}_\alpha={\cal D}_\alpha X^{\underline
a}-i{\cal D}_\alpha\bar\Theta\Gamma^{\underline a}\Theta=0, \quad
{\cal D}_A=({\cal D}_\alpha, {\cal D}_a)=e^{~M}_{A}(z)\partial_M,
\end{equation}
where $e^{~M}_{A}(z)$ is the inverse of the supervielbein matrix
(\ref{e}).

The integrability of (\ref{embflat}) requires
\begin{equation}\label{integ}
\gamma^a_{\alpha\beta}{E}^{~\underline a}_a
=\gamma^a_{\alpha\beta}({\cal D}_a X^{\underline a} -i{\cal
D}_a\bar\Theta\Gamma^{\underline a}\Theta)={\cal
D}_\alpha\bar\Theta\Gamma^{\underline a}{\cal D}_\beta\Theta.
\end{equation}
Using eqs. (\ref{embflat}), (\ref{integ}) and the symmetriesed
gamma--matrix identities in $D=4$
\begin{equation}\label{gammai}
(\Gamma_{\underline{ab}}\Gamma^{\underline
b})_{\{\underline\alpha\underline\beta\underline\gamma\underline\delta\}}=0,
\quad
(\Gamma_{\underline{a}}\Gamma^{\underline
a})_{\{\underline\alpha\underline\beta\underline\gamma\underline\}\delta}=0,
\end{equation}
one can reduce the supermembrane action in the flat target
superspace to the following simple form
\begin{equation}\label{a2d4p}
S=-{i\over{3!}}\int d^3\xi
d^2\eta~det^{-1}(e^{~m}_a)~(\bar\Theta\Theta)~{ E}^{~\underline
a}_a{E}^{~\underline b a}\eta_{\underline a\underline b}+\int
d^3\xi d^2\eta P^\alpha_{\underline a}{E}^{~\underline
a}_{\alpha},
\end{equation}
which resembles the Howe--Tucker--Polyakov term of the action for
the p--branes, though it contains both the Nambu--Goto and the
Wess--Zumino part of the conventional action.

\subsection{The physical gauge}
We now gauge fix all the local symmetries of the supermembrane
action.

The super--Weyl (\ref{superW}) and local $SO(1,2)$ transformations
 allow one to choose the spinor--spinor part of the worldvolume
 supervielbein to be the unit matrix
 $$
 e^{~\mu}_\alpha=\delta^{~\mu}_\alpha.
 $$
 The worldvolume diffeomorphisms are fixed by imposing a physical
 gauge. To this end we split the target space coordinates into the
 ones along and transverse to the membrane
 \begin{equation}\label{split}
 X^{\underline a}=(X^a, X^3(\xi,\eta)), \quad
 \Theta^{\underline\alpha}=(\theta^\alpha,\Psi_\alpha(\xi,\eta))
 \end{equation}
and identify $X^a$ and $\theta^\alpha$ with the superworldvolume
coordinates
\begin{equation}\label{physg}
X^a=\xi^a, \quad \theta^\alpha=\eta^\alpha.
\end{equation}
Note that in this way we also identify 1/2 of space--time
supersymmetry (which shifts $\theta^\alpha$) with global
worldvolume supersymmetry
\begin{equation}\label{susy1}
\delta\theta^\alpha=\delta\eta^\alpha=\epsilon^\alpha_1.
\end{equation}
It is this supersymmetry that remains unbroken.

The spontaneously broken part of space--time supersymmetry is the
one which shifts $\Psi_\alpha(\xi,\eta)$. Thus
$\Psi_\alpha(\xi,\eta)$ is the Goldstone fermion superfield
associated with this symmetry. The form of its transformation
under spontaneously broken supersymmetry is dictated by the
requirement that the physical gauge conditions remain invariant
under this symmetry. This transformation is easily found to be
\begin{equation}\label{susy2}
\delta\Psi_\alpha=\epsilon^2_\alpha+i(\bar\epsilon^2\gamma^m\Psi)\partial_m\Psi_\alpha.
\end{equation}
We observe that broken supersymmetry is nonlinearly realized in
the transformations of the Goldstone fermion.

The superfield $X^3(\xi,\eta)$ is the Goldstone scalar associated
with spontaneously broken translations transverse to the membrane.
$\Psi_\alpha(\xi,\eta)$ and $X^3(\xi,\eta)$ are not independent.
The transverse ($\underline a=3$) component of the superembedding
condition (\ref{embflat}) relates them as follows
\begin{equation}\label{tran}
E^3_\alpha=0 \quad \Rightarrow \quad {\cal D}_\alpha
\Phi(z)=2i\Psi_\alpha(z),
\end{equation}
where
\begin{equation}\label{X3}
\Phi=X^3+i\eta^\alpha\Psi_\alpha, \quad {\cal
D}_\alpha=\partial_\alpha+e^{~m}_\alpha(z)\partial_m.
\end{equation}
And the part of the superembedding condition parallel to the
brane expresses the remaining independent supervielbein components
$e^{~m}_\alpha(z)$ in terms of $\Psi_\alpha$. $e^{~m}_\alpha(z)$ thus
become induced by the superembedding, as we discussed in the
Introduction
$$
E^m_\alpha =0 \quad \Rightarrow \quad
e^{~m}_\alpha=i\gamma^m_{\alpha\beta}\eta^\beta+iD_\alpha\bar\Psi\gamma^m\Psi-
D_\alpha\bar\Psi\gamma^b\Psi
\partial_b\bar\Psi\gamma^m\Psi
$$
\begin{equation}\label{epsi}
=i\gamma^m_{\alpha\beta}\eta^\beta+iD_\alpha\bar\Psi\gamma^b\Psi
(\delta_b^m+i\partial_b\bar\Psi\gamma^m\Psi),
\end{equation}
where
\begin{equation}\label{Da}
D_\alpha={\partial\over{\partial\eta^\alpha}}+i\eta^\beta\gamma^a_{\beta\alpha}
{\partial\over{\partial\xi^a}}, \quad
\{D_\alpha,D_\beta\}=2i\gamma^a_{\alpha\beta}
{\partial\over{\partial\xi^a}}
\end{equation}
are covariant derivatives in a flat $N=1$, $d=3$ superspace.

We have thus shown that in the physical gauge the fluctuations of
the $N=1$, $D=4$ supermembrane are described by a single
unconstrained scalar superfield $\Phi(\xi,\eta)$ , the Goldstone
boson associated with broken translations in the direction
transverse to the membrane. The Goldstone spinor
$\Psi_\alpha(\xi,\eta)$ is expressed through $\Phi(\xi,\eta)$ and
its derivatives in a highly nonlinear way. This is an example of
a so called inverse Higgs effect \cite{ivanov}. The effect is that
under certain covariant conditions the number of Goldstone fields
gets reduced by making some of them dependent on the others. We
should note that the inverse Higgs effect is only part of the
superembedding condition, which is more general, and in
particular also ensures the superworldvolume geometry to be
induced by the embedding, as we have seen above.

We can now substitute the expressions, which we have found in the
physical gauge, into the covariant supermembrane action and upon
some calculations get the nonlinear Goldstone superfield action
in the following form
\begin{equation}\label{aphys1}
S=iT\int d^3\xi d^2\eta~{\Psi^2\over{1-{1\over 4}D^2\Psi^2}}+T\int
d^3\xi\cdot 1\, ,
\end{equation}
where the second term in (\ref{aphys1}) is the membrane ground
state ($\Psi_\alpha=0$) energy,
$D^2=D^\alpha D_\alpha$,
$\Psi^2=\Psi^\alpha\Psi_\alpha$ and the Goldstone fermion
$\Psi_\alpha$ depends on the Goldstone scalar $\Phi$
(\ref{tran}), i.e. $\Psi_\alpha=D_\alpha\Phi~ +
~(nonlinear~terms)$. Though the explicit expression of $\Psi$ in
terms of $\Phi$ is rather involved, it is nevertheless possible
to get from (\ref{aphys1}) the action for the independent
superfield
$\Phi(\xi,\eta)$ \cite{pst}
\begin{equation}\label{aphys2}
S=-{iT\over 2}\int d^3\xi d^2\eta~{D^\alpha\Phi
D_\alpha\Phi\over{1-{1\over
8}(D^2\Phi)^2}+\sqrt{1+\partial_a\Phi\partial^a\Phi(1-{1\over
16}(D^2\Phi)^2)}}+T\int d^3\xi\cdot 1\, .
\end{equation}
Eq. (\ref{aphys2}) is the superfield form of the gauge fixed
component action for the $N=1$, $D=4$ supermembrane obtained in
\cite{achu}.

On the other hand, the action (\ref{aphys1}) can be reduced to
the Goldstone superfield action constructed in \cite{ik1} with the
use of the method of a `linear' realization of spontaneously
broken supersymmetry. The action of \cite{ik1} describes the
dynamics of a Goldstone scalar superfield $\rho(\xi,\eta)$
different from
$\Phi(\xi,\eta)$ of (\ref{aphys2}). The explicit relation between
$\rho(\xi,\eta)$ and $\Phi(\xi,\eta)$ has not been presented in the
literature. However, there exists \cite{ik1} the expression of
$\Psi_\alpha(\xi,\eta)$ in terms of $\rho(\xi,\eta)$
\begin{equation}\label{psizeta}
\Psi_\alpha={\zeta_\alpha\over{ 1+D^2{\cal F}}}, \quad
\zeta_\alpha=D_\alpha \rho(\xi,\eta),
\end{equation}
where
\begin{equation}\label{tilphi}
{\cal F}={1\over 2} {\zeta^2\over{1+\sqrt{1+D^2\zeta^2}}}.
\end{equation}
Substituting (\ref{psizeta}) into (\ref{aphys1}) we get the
action for $\rho(\xi,\eta)$ found in \cite{ik1}
$$
S=2Ti\int d^3\xi d^2\eta
{(D\rho)^2\over{1+\sqrt{1+D^2(D\rho)^2}}}+T\int d^3\xi \cdot 1\,.
$$

\section{Conclusion}
Starting from a covariant superembedding formulation of
supermembrane dynamics in $N=1$, $D=4$ superspace and having
gauge fixed the superworldvolume local symmetries we have got an
effective nonlinear superfield theory on the brane
superworldvolume which exhibits partial supersymmetry breaking.
We have also demonstrated how the supermembrane is related to a
model of partial breaking of $N=2$, $d=3$ supersymmetry discussed
in \cite{ik1}. The superembedding approach has provided us with a
systematic way of doing this.

As a generalization of these results, it should be possible to get
a $D=4$ Dirac--Born--Infeld action with partially broken $N=2$
supersymmetry from a superembedding formulation of a
space--filling D3--brane in $N=2$, $D=4$ superspace, and thus to
establish the direct relationship between the D3--brane and the
Goldstone superfield actions of \cite{bg2,rt,fer}.

One may also address a problem of whether the conditions and
symmetries associated with superembeddings may allow one to
overcome ambiguities in the construction of the non--Abelian
generalization of the Dirac--Born--Infeld theory, and hopefully
to make a progress in the target space covariant description of
the system of $N$ coincident D--branes and its embedding into
target {\it super}space. In a conventional ``kappa--symmetric"
approach a study of this problem has been undertaken in \cite{em}.

The methods of superembeddings can also be used to study the
possibility of obtaining a simpler form of superconformal field
theory actions on the Anti--de--Sitter boundary, as well as
studying effects of local supersymmetry breaking in supergravity
theories.
\\

\noindent
{\bf Acknowledgements.} This work was partially supported by the
European Commission TMR Programme ERBFMPX-CT96-0045 and the RTN
Programme HPRN-CT-2000-00131 to which the authors are associated.
D.S. also acknowledges partial support from the Grant N
2.51.1/52-F5/1795-98 of the Ukrainian Ministry of Science and
Technology.


\end{document}